\documentclass[12pt]{article}
\usepackage{mathtext}
\usepackage{graphicx}
\usepackage{amsfonts}
\usepackage{amssymb}
\usepackage{amsmath}
\numberwithin{equation}{section}

\newcommand{\e}{\epsilon}

\newcommand{\la}{\lambda}

\begin{document}
\title{On the physical limit of quantum computing}

\author{Y.I.Ozhigov\thanks{ozhigov@cs.msu.su}\\
Moscow State University of M.V.Lomonosov \\
VMK Faculty,\\
Institute of physics and technology RAS}
\maketitle

\begin{abstract}
Experimental attempts to implement quantum speedup of computations over the past 30 years have yielded a negative result, despite the absence of physical laws prohibiting such speedup. The article formulates the limitation of quantum formalism in the form of uncertainty ''the complexity of the system - the accuracy of its description at the quantum level'', and provides arguments in favor of its physical status. An experiment to determine this constant through Grover's algorithm is described. Rough estimates on the constant of this ratio are given, based on the possibility of applying the quantum theory to two processes: the emission of a photon by a Rubidium atom and the decay of an unstable isotope of Helium 6. This ratio explicitly prohibits the physical implementation of scalable fast quantum computations, but leaves the possibility of modeling the dynamics of real systems on a quantum computer, the only advantage of which is the use of quantum nonlocality.
\end{abstract}
\newpage
\section{Introduction}

The analysis of infinitesimals is the ''sacred cow'' of exact natural science, despite the fact that the pursuit to the limit itself is a simplification of the real picture. For example, any continuous medium consists of particles of finite size, so the traditional description of such media through differential equations is already an assumption. The hypnosis of infinitesimals was overcome for the first time in 200 years after Newton by quantum mechanics, which postulated the quantization of the action in the form of Planck constant $\hbar\approx 10^{-27}\ erg\ sec$, as well as the quantization of the energies of the stationary electronic states in atoms.  

This great physical discipline turned out to be closely related to the science of computing - through the quantum computer project, so that the discrete ideology of computing actually began to play the role of the mathematical apparatus of the quantum theory of complex systems\footnote{the opposite is also true: the mathematical problems of the algorithm theory of the type $P=NP(?) $ can apparently be satisfactorily solved only by physics.}. In this paper, we formulate a restriction on quantum formalism and quantum computations in terms of the theory of complexity, based on almost 30-year history of experiments on a quantum computer. We also propose the experimental method for determining the parameters of this restriction - as a constant in the uncertainty relation ''complexity of the system - the accuracy of its description at the quantum level''. This explicitly prohibits fast quantum computing, but leaves open the development of quantum computer as a device modeling real processes.

Quantum mechanics, as a universal means of accurate description of nature has a single, but deadly drawback - algorithmic complexity. The memory of the computer modeling quantum evolutions generally grows as an exponent of the number of real particles in the system under consideration, which makes such modeling impossible even for relatively simple systems such as molecules of several atoms. This computational complexity was the reason for the appearance of the quantum computer project, a new look at which this article presents.

At the physical level, any computation is, by its nature, quantum, and therefore the analysis of the quantum computer is an analysis of our capabilities to simulate the reality as a whole. At the end of the 20th century, a tempting way to solve the problem of modeling complex systems was proposed: a quantum computer (see the pioneering work of R. Feynman \cite{Fe3}), it was also proved that such modeling does not contradict the standard quantum theory, and requires memory growing linearly with the increase in the number of real particles in the system, however, with a quadratic time dilation compared to the real process (see works \cite{Za}). Almost simultaneously it was discovered an amazing property of quantum computations - their ability to solve mathematical problems of the search type, with the option of a ''black box'' faster than any classical computer (\cite{Sh}, \cite{Gr}), and also proved that the quantum speedup holds only for a very limited range of problems (\cite{BB_},\ \cite{BB__},\ \cite{Oz}), in particular, hope to speed up with it the GMSP (generic machine simulation problem - see the definition in \cite{GMSP}) is empty (\cite{Oz3}).

Experiments on the practical construction of a quantum computer, rapidly started in the early 90-ies of the last century, continue now. They did not give confidence in the scaling of the Feynman quantum computer in its abstract-radical form, assuming the physical existence of a Hilbert space with the number of dimensions $2^n$, where $n$ is the hundreds, whereas it is this hypothesis put forward and defended by most physicists that underlies fast quantum algorithms.

A brief history of these experiments is as follows. The work began with NMR structures (one of the last works in this area \cite{NMR}), for the main reason: the nuclear spin states live hours that is very important for fragile quantum states. The further modifications were proposed, like a combination of charge states of electrons with spin states (see, for example, \cite{LV}). 
Quantum superconducting elements (Josephson junctions and squids: 
\cite{DS},\ \cite{PG},\ \cite{AG}) are attractive for their mesoscopic dimensions. 
Photonic gates (for example, \cite{KLM},\ \cite{Az}) have been intensively developed in the last 10 years - their advantage is the proximity to the instrument park of quantum cryptography. 
Hybrid computing: cold ions in Paul traps (\cite{LBM}), optical-mechanical systems (e.g. \cite{Tsu}). 
Computing systems of the topological type, for example the anions (\cite{Ki}), occurring in the subspaces, a-priori free from decoherence; they either rely on not yet studied particle types, or are facing difficulties in the operations that are the greater the protection against decoherence is more reliable.

There is also a mathematical method of dealing with decoherence - quantum error correction codes (one of the pioneer works is \cite{ShC}). These codes start with only a few thousand qubits; their application requires a quantum computer to work well with that number of qubits without correction codes, and it is this barrier that may fall into the critical domain of the restriction on formalism itself, which is discussed below.

It can be concluded that the decoherence - spontaneous destruction of quantum states necessary for the regulated operation of a quantum computer, is not a technical problem. The traditional interpretation of decoherence as a ''harmful influence of the environment'' (see the concept of an open quantum system \cite{BP}) can hardly be considered as satisfactory, since this phenomenon invariably destroys quantum computing regardless of technology. In addition, in quantum mechanics itself, the ''influence of the environment'' is a methodologically dark place, since the most harmful form of this influence, the so-called measurement, implies the existence of an observer - an object that obeys classical laws and is fundamentally separated from the quantum system.

Sensitivity to decoherence is a special property of quantum states, which does not coincide with the degree of their entanglement. Thousands of atoms can be in entangled states (for example, acoustic vibrations of atoms in diamond - see \cite{LSS}, \cite{LBM}), but these states do not provide fast algorithms. Whereas the states that are necessary for quantum speedup are surprisingly very sensitive to decoherence.

These arguments, along with the apparent change that quantum computing brings to the theory of algorithmic complexity, suggest a deep internal connection between the algorithmic description of nature and quantum mechanics. In particular, they raise the suspicion that the algorithmic description of real evolutions, with all its limitations, can be an independent form of physical laws.

In quantum mechanics, there is already a thesis that the form of describing reality is a property of reality itself - this revolutionary thesis for the beginning of the 20th century is confirmed by the fundamental Bohr-Heisenberg uncertainty relation: $\delta x\ \delta p=\hbar$, according to which the better the momentum can be known, the worse the coordinate can be known and vice versa, as well as a similar ''energy - time'' ratio. 

If we assume that there is a law prohibiting fast quantum computations, its simplest form is the uncertainty relation ''the complexity of the system - the accuracy of its quantum description''. We present arguments in favor of such a relation, and show how one can experimentally determine a constant in this relation analogous to Planck constant $\hbar$. This experiment is based on the well-known Grover algorithm for solving the NP - complete search problem (\cite{Gr}), and it is feasible to carry it out on the available prototypes of bounded quantum computers if the given constant is not much greater than $10^9$.

\section{Feynman quantum computer and quantum computing}

Feynman quantum computer is a universal model of complex systems at the quantum level. Its idea stems from the structure of quantum theory itself. It has a classical part, the purpose of which is to control the evolution of the quantum part with the help of classical fields. The latter consists of $n$ qubits (the number of which can be increased), each of the $2^n$ classical states $|j\rangle$ of which we identify with a binary string. A quantum state of the quantum part has the form:

\begin{equation}
\label{state}
|\Psi\rangle=\sum\limits_{\j\in J}\la_j|j\rangle.
\end{equation}

In this case, the quantum algorithm indicates the classical control action on these qubits in the form of a sequence of simple unitary operations - quantum gates. The corresponding evolution of the quantum part $|\psi_0\rangle\rightarrow |\psi_1\rangle\rightarrow ... \rightarrow |\psi_t\rangle$ is called quantum computation. The adiabatic model of computation (see, for example, \cite{Ag}) is quite similar in theory, only there instead of gates the law of slow change of the Hamiltonian of the quantum system is used.
In any case, the result of quantum computation is either a finite state $|\psi_t\rangle$ of the qubit system, or is obtained from it by measurement.

Note that the amplitudes in \eqref{state} are physically dimensionless. When modeling real processes on a quantum computer (for example, in the spirit of \cite{Za}), one should multiply these amplitudes by a factor $d^{-1/2}$, where $d$ is the physical dimension of the classical basic states $|j\rangle$; for example, for a three-dimensional particle $d=cm^3$.

To elucidate the limits of application of quantum theory in computing, we address to quantum Grover search algorithm GSA (see \cite{Gr}), which finds the solution of the equation
\begin{equation}
f(x)=1
\label{G}
\end{equation}
where $f$ is a Boolean function of $n$ variables given by a scheme of classical functional elements. GSA finds one of $l$ solutions of the equation \eqref{G} in $O(\sqrt{N/l})$ steps, where $N=2^n$, while the classical computer requires at least $\Omega(N/l)$ steps. The quantum computation time thus has the order of the square root of classical time. In what follows we consider the case $l=1$, by $x_t$ we denote the only solution of \eqref{G}.

The fast quantum computation leads to the goal by concentrating the value of amplitudes $\la_j$ of all basic states $|j\rangle$ in \eqref{state} on the target state $|x_t\rangle$ to be found. This process can be made much faster than the classical computation of $x_t$ provided that the set $J$ in \eqref{state} from which the amplitude is collected is exponentially large, not achievable for any classical computer.

 We consider the action of oracle $f_q:\ |x,y\rangle\rightarrow |x,y\oplus f(x)\rangle$ on the state $|x\rangle|\tilde 0\rangle$, where the state of ancilla is $|\tilde 0\rangle=\frac{1}{\sqrt 2} (|0\rangle-|1\rangle)$: it will give the same state if $x\neq x_t$, and change sign if $x=x_t$. Thus, the quantum oracle acts as a mirror reflection $I_{x_t}$ of the entire state space relative to the subspace orthogonal to $|x_t\rangle$.

It is also easy to construct an auxiliary operator $I_{\tilde O}$ that mirrors the state space with respect to the subspace orthogonal to the state $|\tilde O\rangle=\frac{1}{\sqrt N}\sum\limits_{j=0}^{N-1}|j\rangle$, where $N=2^n$. Having these two mirror reflections, we construct the operator $G=-I_{\tilde O}I_{x_t}$, which on the two-dimensional real plane generated by the vectors $|x_t\rangle$ and $|\tilde O\rangle$ acts as a rotation by the angle $2\ arcsin\frac{1}{\sqrt N}$, and therefore applying $ [\pi \sqrt N/4]$ times\footnote{Square brackets here denote the integer part of the number.} the operator $G$ starting with $|\tilde O\rangle$, we get $|x_t\rangle$ with very high accuracy. It is seen that the complexity of such an algorithm for finding the solution of the equation \eqref{G} has the  order of the root of the classical time $N$ of the solution of this problem: $t_{quant}=O(\sqrt t_{class})$. This is quantum amplitude amplification.

Quantum speedup of classical computation is a rare phenomenon. It should be considered, first of all, as a criterion for building a real quantum computer, and not all its ersatz. In Copenhagen quantum mechanics there is no ban on its existence (see \cite{Va}), and therefore its construction is a radical test of quantum theory in the area where it has never been tested - in the field of complex systems. Here, the only form of verification will be the implementation of a fast quantum algorithm, for example GSA.

The normalization condition of the wave function $\sum\limits_{j=0}^{N-1}|\la_j|^2=1$ means that quantum speedup by GSA  is possible only when the amplitudes of $\la_j$ can be exponentially small with the growth of $n$. But in this case, the direct calculation of these amplitudes according to the Born's rule - with the help of a set of statistics, becomes impossible. And the only way to verify quantum theory is the implementation of GSA. This is problematic: here we are dealing with the limits of applicability of quantum theory and therefore it is necessary to consider its limitation  - the uncertainty relation ''accuracy of the description of the state vector - quantum complexity of the system'', and to interpret experiments on the quantum computer as finding the constant in this ratio.

\section{Constructive representation of the wave function of a multiparticle system}

Advances in quantum theory are based on the effective calculation of the amplitudes of $\la_j$, which for real experiments is performed on computers. In computer representation, amplitudes are always quantized, that is, they have the form:
\begin{equation}
\label{lambda}
\la_j=(k_j+il_j)\epsilon,
\end{equation}
 where $\epsilon$ is a small nonzero value - amplitude quantum, and $k_j,\ l_j$ are integer numbers. This representation of ampitudes follows from the linearity of quantum theory. It also requires an appropriate choice of classical basic states, but due to the smallness of $\epsilon$ this does not lead to any revision of the experimentally confirmed part of the quantum theory, but will only affect the scaling of quantum computer.

So, in our algorithmic approach, there are only states of the form 
\begin{equation}
\label{state1}
|\Psi\rangle=\sum\limits_{j\in J}\la_j|j\rangle,
\end{equation}
where the amplitudes have the form \eqref{lambda}. So the summation in \eqref{state1} extends to no more than $1/\epsilon^2$ terms from the set $J=\{ 0,1,...,N-1\},\ N=2^n$. 

The accuracy of the predictions of quantum theory is associated with the numerical computations, which involve only numbers of the form \eqref{lambda} with a certain minimum allowable for storage in computer memory $\e$. In this case, computer operations on such states implement any numerical methods (Euler, Runge-Kutta, etc.), but do not lead beyond such a discrete class of states. Therefore,  the discrete constraint we introduced will not lead to any revision of the results of quantum mechanics. However, it will radically change our interpretation of quantum computer and the meaning of experiments on its creation, as well as the form, in which quantum theory may be expanded to complex systems.

\section{About the size of the amplitude quantum}

The natural question: what is the size of the amplitude quantum, has no exact answer, just as the question about the limit of accuracy of measurements. On one hand, it is known that for very simple systems, for example, for an electron in a hydrogen atom, $\epsilon\approx 0$ that is, this quantum is simply negligible and does not play any role: the wave functions of stationary states are determined very accurately. Here a huge number of equally ''prepared'' hydrogen atoms, which can be measured in parallel mode plays a role, so that the collected statistics will perfectly correspond to the analytically found wave function.

It could be assumed that the amplitude quantum and the number of particles in the system in question are inversely proportional or something like that, but this also would not be true. A set of interacting with each other harmonic oscillators can be arbitrarily large, but the wave function of such a system can be determined very accurately, since its basic states can be overridden by a quasi-particle representation, so that the resulting quasi-particles will not interact with each other, and therefore the accuracy of the definition of the wave function of the entire system will coincide with the accuracy of its definition for a single harmonic oscillator, that is, can be very large.

Let $M$ be the set of $n$ qubits of the system under consideration. A state $|\Psi\rangle$ of these qubits is called nonentangled if there is such a partition of $M=M_1\cup M_2$ into two disjoint non-empty sets and the state $|\Psi_1\rangle,\ |\Psi_2\rangle$ on these sets such that $|\Psi\rangle=|\Psi_1\rangle\otimes |\Psi_2\rangle$. Otherwise, the state of $|\Psi\rangle$ is called entangled. The complexity of the state $|\Psi\rangle$ on the set $M$ is the size in qubits of maximum its entangled tensor divider, i.e. the maximum of natural numbers $s$ such that there exists a subset of $M_1\subseteq M$ and the states $|\Psi_1\rangle,\ |\Psi_2\rangle$ on $M_1$ and $M-M_1$, respectively, such that $|\Psi\rangle=|\Psi_1\rangle\otimes |\Psi_2\rangle$, $M_1$ contains $s$ elements and $|\Psi_1\rangle$ is entangled. This state $|\Psi_1\rangle$ is called the quantum kernel of the state $|\Psi\rangle$, and the corresponding set $M_1$ is the kernel carrier.

There can be several kernels, since the maximum number of $s$ from the definition can correspond to different kernel carriers. Naturally, this definition may depend on very small amplitudes, so that the complex state can be very close to simple. However, if we consider only state, which amplitudes $\la_j$ have the form \eqref{lambda}, this closeness will limited by the value of $\epsilon$. From further it will be clear that it is impossible to direct $\epsilon$ to zero for complex systems, and therefore the complexity is determined in this way correctly. We will denote the complexity of the state $|\Psi\rangle$ by $C(\Psi)$.

Let $\tau\in S_N$ be the permutations of the basis vectors of the main state space corresponding to the set of qubits $M$. Then the naturally defined state $\tau |\Phi\rangle$ is called a quasi-particle representation of the state $|\Psi\rangle$. For example, for a set of $n$ harmonic oscillators each of their basic states has the form $ (q_1,q_2,...,q_n)$ and the Fourier transform over this sequence of the form $Q_k=\alpha\sum\limits_jq_je^{-\beta\ ikj}$ means the transition to the description of the same system of oscillators through phonons - quasiparticles with new coordinates $Q_k$. 
The other example: generalized GHZ - state of the form $\frac{1}{\sqrt 2}(|00...0\rangle+|11...1\rangle)$ in which all $n$ qubits are entangled, but it can be reduced to the completely nonentangled state by successive operations of $CNOT$, which are permutations of the basic states.

The absolute complexity $A(\Psi)$ of the  state $|\Psi\rangle$ is the minimum of the complexities of all its quasi-particle representations. Formally:

\begin{equation}
\label{abs_complexity}
A(\Psi)=min_{\tau\in S_N}C(\tau |\Psi\rangle).
\end{equation}
 
Absolute complexity is the number of qubits required to represent the quantum kernel of a given state. The state space in which this kernel lives has thus a dimension of $2^{A(\Psi)}$, which we will further denote by $N$.

We assume that for the state \eqref{state1} the amplitudes of its components always have the form \eqref{lambda} for some $\epsilon>0$.

It is the same as the amplitude will take only four values, $\pm\epsilon,\ \pm i\epsilon$, and the states from the set $J$ in the expansion \eqref{state1} take some distribution in the classical space ${\cal K}$ of configurations of the system, so if we want to calculate the "wave function" in the usual sense of the term, we must sum the amplitudes for small segments $\delta_k{\cal K}$ of the space ${\cal K}$ so four types of "amplitude quanta", being added for all states in $J\cap \delta_k{\cal K}$, give the value of the "wave function" in the center of the segment $\delta_k{\cal K}$. 

We assume that determining of the spatial position of any basic state $j\in J$ requires little memory, as well as the notation of of amplitudes since the notation of a number is logarithmic in length. So the maximum value $Q$ of the number $N$ of the basis states in the kernel of any quantum system is a constant independent of the state $ |\Psi\rangle$. Then the value $\epsilon=1/\sqrt{Q}$ will also be a constant, and dimensionless, since the dimension of the physical quantity will refer to the basic states $|j\rangle$ in \eqref{state1}.

The representation of the wave function with $N$ basic states is achieved in the only case - when all amplitudes are modulo $\epsilon$. If the basic states are less than $Q$, it means that the quantum amplitude is summed on segments $\delta_k{\cal K}$, and then we have the uncertainty relation ''complexity-accuracy'' of the form
\begin{equation}
N\ log_2(1/\varepsilon)\leq Q
\label{uncertainty}
\end{equation}
where $\varepsilon$ is the accuracy of representing the amplitudes of a traditional ''wave function'' for a set of $N$ points in space ${\cal K}$. So, for values $N\ll Q$, the accuracy of the representation of the traditional ''wave function'' will be so high that it will be impossible to distinguish it from the analytically obtained expression; for simple systems this is the case.

\section{Experimental determination of the constant $Q$}

The constructive dimensionless constant $Q$ has a physical nature, and therefore is subject to experimental search. To do this, it is necessary to distribute the amplitude over a very large number of classical states of some ensemble - on its kernel, so that the presence of a quantum of amplitude $\epsilon$ would lead to a gross deviation from coherent dynamics, which could be recorded in the experiment. If we deal with well-studied physical ensembles, it is very difficult to do so, since they are amenable to study by standard means precisely because their absolute complexity is small and therefore the accuracy of determining the amplitude can be very high.

The most reliable way is to implement Grover search algorithm - GSA  (see \cite{Gr}) for as many $n$ qubits as possible. GSA is the sequentional application of Grover operator $G$. We have to reproduce the computation from this algorithm to find the root of the equation

\begin{equation}
\label{grov}
f(x)=1, 
\end{equation}
where $f$ is a Boolean function of $n$ variables, such that this equation has the unique solution $x_t$. 
We note that after $s$ applications of $G$ the state with the high accuracy will have the form $cos(2s/\sqrt{N})|\tilde 0\rangle+sin(2s/\sqrt{N})|x_t\rangle$, where $|\tilde 0\rangle=\frac{1}{\sqrt N}\sum\limits_{j=0}^{N-1}|j\rangle$. This typical form, when the amplitude of a single state exceeds the amplitude of all other states, in which it is the same, will remain at any permutation $\tau$ of the basic vectors, so starting from the first step all $n$ qubits will form a single quantum kernel, and the maximum number of $n$ qubits, for which GSA will give the correct answer, will give the value $Q=2^n$. Here we assume that GSA works if at least for the first application of $G$ the exceeding the amplitude of the target state over all others is obtained; to fix this, it is necessary, of course, to make numerous measurements of the result of one such application, which with a large value of $Q$ can become a problem. Anyway, the experiment can estimate $Q$ from below if you try to bring the GSA to the end by making $[\pi\sqrt{N}/4]$ applications of $G$.

\section{Physical sense of $Q$}

We give estimates of the constant $Q$ in connection with the applicability of quantum theory to processes of different nature. These estimates are very rough, but they allow us to link the difficult question of the applicability of quantum theory to various complex processes with a single experiment: the implementation of GSA.

Consider two processes: the transition of electron states in an atom $Rb^{85}$ and the decay of an unstable nucleus $He^6$. The first process is described by quantum electrodynamics quite accurately, a complete quantum description of the second is not yet available.

We will proceed from the criterion of accurate drawing of the wave function, when each step of its computer description requires one new basis state. This follows from the speed of the quantum walk, in which the wave front propagates at a linear speed (as opposed to the classical walk, in which the speed is proportional to the square root of time). Let $t$ be the total time of the process, $dt$ be the step of the computer description of this process in time, then the number of basic states required for ''accurate drawing'' of the process is $N=t/dt$. The values of $t$ are determined experimentally, and $dt$ is derived from the energy - time uncertainty relation.

For Rabi oscillation of the rubidium atom occurring with the emission of a photon with a wavelength of approximately $1.4\ cm$ we have: $\omega \approx 10^{10}\ sek^{-1}$, $E_{QED}=\hbar\omega\approx 10^{-17}$, $dt\approx\hbar/E_{QED}=10^{-10}$. Given the time of Rabi oscillation $t\approx 10^{-6}\ sek$, we get $N=t / dt\approx 10^{4}$. Thus, if this process can be well drawn on the basis of quantum theory, then $Q\geq 10^{4}\approx 2^{13}$ and GSA should work on about 13 qubits, which seems quite real.

Now consider the decay of the nucleus of the Helium isotope: $He^6\rightarrow He^5 + n \rightarrow He^4 + 2n$ (in this rough approximation, we consider only nucleons). The characteristic energy value will be about $10\ Mev\approx 10^{-5}\ erg$, and the energy-time uncertainty relation will give $dt\approx 10^{-22}\ sec$. The whole process takes about $1.6 \ sec$, from where $N=t / dt\approx 10^{22}\approx 2^{73}$, and if quantum mechanics can be continued to nuclear processes like $He^6$ isotope decay to a stable isotope $He^4$, GSA should work well already at $73$ qubits.

The decay of $He^6$ isotope is a complex process. It is possible to consider only its last stage when one neutron is split off the unstable nucleus $He^5$ and flies away giving the stable $He^4$. It takes approximately $10^{-11}\ sec$. For it, estimates similar to the above will give about $36$ qubits of a reliable implementation of GSA, which is less realistic than the electrodynamic case, but the corresponding value of $Q\approx 2^{36}$ can already be verified by experiments on GSA.

 Thus, the acceptance of the accuracy - complexity uncertainty relation hypothesis directly links the question of the applicability of quantum theory to real microprocesses and the implementation of GSA. The implementation of GSA thus becomes a central issue of quantum theory and the theory of complex systems as such.

\section{Conclusion}

We proposed the uncertainty relation ''the accuracy of the description of the wave function - the complexity of the quantum system'', the constant of which can be found by analyzing the behavior of the state of the quantum computer that implements Grover search algorithm. The number of qubits for which this algorithm will operate normally, and the accuracy with which the wave function can be practically found, directly determines this constant.

In the standard formulation of quantum theory, this constant is equal to infinity. This makes it impossible to directly find the wave function of the system of many particles by means of statistical analysis. Thus, in the standard formulation, the quantum theory of many bodies can be verified only by implementing the fast quantum Grover algorithm in its scalable form, which is doubtful in the light of the current state of the quantum computer experiments.

Time and energy of specific processes allow to estimate the value of this constant. For example, the implementation of GSA for more than 70 qubits is needed in order to ensure that nuclear processes can be described by the methods of quantum theory as confidently as the motion of an electron in the hydrogen atom. The possibility of a very accurate quantum description of the electron dynamics in an atom implies the possibility of implementing GSA for about 12 qubits. This determines the importance of experiments on the implementation of this algorithm.

This uncertainty relation actually gives a limit of applicability of quantum theory in terms of algorithmic complexity, and this limit is quite achievable in experiments. 

\section{Acknowledgements}

The work is supported by the Russian Foundation for Basic Research, grant a-18-01-00695.

\end{document}